\documentclass[aps,prl,reprint,groupedaddress]{revtex4-1}
\usepackage{graphicx} 
\usepackage{dcolumn} 
\usepackage{bm}
\usepackage{hyperref}

\begin{document}

\title{Nuclear in-medium effects of strange particles in proton-nucleus collisions}
\author{Zhao-Qing Feng$^{1,2}$}
\email{fengzhq@impcas.ac.cn}

\author{Wen-Jie Xie$^{1}$}

\author{Gen-Ming Jin$^{1}$}

\affiliation{$^{1}$Institute of Modern Physics, Chinese Academy of Sciences, Lanzhou 730000, People's Republic of China            \\
$^{2}$State Key Laboratory of Theoretical Physics and Kavli Institute for Theoretical Physics China, Chinese Academy of Sciences, Beijing 100190, People's Republic of China}

\date{\today}

\begin{abstract}
Dynamics of strange particles produced in proton induced nuclear reactions near threshold energies has been investigated within the Lanzhou quantum molecular dynamics (LQMD) transport model. The in-medium modifications on particle production in dense nuclear matter are considered through corrections on the elementary cross sections via the effective mass and the mean-field potentials. It is found that the attractive antikaon-nucleon potential enhances the subthreshold $\overline{K}$ production and also influences the structure of inclusive spectra. The strangeness production is strongly suppressed in proton induced reactions in comparison to heavy-ion collisions. The kaon-nucleon and antikaon-nucleon potentials change the structures of rapidity and transverse momentum distributions, and also the inclusive spectra. The measured K$^{-}$/K$^{+}$ ratios in collisions of $^{12}$C+$^{12}$C, $^{197}$Au+$^{197}$Au, p+$^{12}$C, and p+$^{197}$Au from KaoS collaboration have been well explained with inclusion of the in-medium potentials.

\begin{description}
\item[PACS number(s)]
25.40.-h, 21.65.Ef, 21.65.Jk
\end{description}
\end{abstract}

\maketitle

Properties of hadrons in nuclear medium is interest in studying the QCD structure in dense matter, in particular related to the chiral symmetry restoration, properties of hypernucleus etc \cite{Gi95,Fr07}. High energy heavy-ion collisions in terrestrial laboratory provide the unique possibility to study both the equation of state and the in-medium properties of hadrons in dense nuclear matter. It was found that the K$^{-}$/K$^{+}$ ratio is enhanced in heavy-ion collisions in comparison with proton-proton collisions \cite{Ba97,Me00,Fo03}. The increase of K$^{-}$ mesons in heavy-ion collisions is partly caused from strangeness exchange reactions ($\pi Y \rightarrow K^{-}N$, $YN \rightarrow K^{-}NN$), and also from the attractive antikaon-nucleon potentials in dense nuclear medium. The measured kaon and antikaon yields can be well reproduced by transport models with taking into account a density dependent mean-field potential \cite{Li97,Ca99,Fu06,Ha12}. The extraction of the in-medium properties of strange particles from heavy-ion collisions is complicated, where the nuclear density varies in the evolution of nucleus-nucleus collisions. To avoid the uncertainties of the baryon densities on strange particle production, one can investigate proton-nucleus collisions where the nuclear density is definite around the saturation density.

In this work, the Lanzhou quantum molecular dynamics (LQMD) model has been used to investigate the proton induced nuclear reactions, in which the dynamics of resonances, hyperons and mesons is described via hadron-hadron collisions, decays of resonances and mean-field potentials \cite{Fe11,Fe13}. The evolutions of baryons (nucleons, resonances and hyperons) and mesons ($\pi$, $K$, $\eta$, $\overline{K}$, etc.) in the course of proton-nucleus collisions are governed by Hamilton's equations of motion. A Skyrme-type momentum dependent interaction has been used in the evaluation of the mean-field potentials for nucleons and resonances. The hyperon mean-field potential is constructed on the basis of the light-quark counting rule, in which the nucleon self-energies are computed from the relativistic mean-field model.

The kaon and anti-kaon energies in the nuclear medium distinguish isospin effects based on the chiral Lagrangian approach as \cite{Fe13,Ka86,Sc97}
\begin{eqnarray}
\omega_{K}(\textbf{p}_{i},\rho_{i})= && \left[m_{K}^{2}+\textbf{p}_{i}^{2}-a_{K}\rho_{i}^{S}
-\tau_{3}c_{K}\rho_{i3}^{S}+(b_{K}\rho_{i}+\tau_{3}d_{K}\rho_{i3})^{2}\right]^{1/2}
\nonumber \\
&& +b_{K}\rho_{i}+\tau_{3}d_{K}\rho_{i3}
\end{eqnarray}
and
\begin{eqnarray}
\omega_{\overline{K}}(\textbf{p}_{i},\rho_{i})= && \left[m_{\overline{K}}^{2}+\textbf{p}_{i}^{2}-a_{\overline{K}}\rho_{i}^{S}
-\tau_{3}c_{K}\rho_{i3}^{S}+(b_{K}\rho_{i}+\tau_{3}d_{K}\rho_{i3})^{2}\right]^{1/2}
\nonumber \\
&& -b_{K}\rho_{i}-\tau_{3}d_{K}\rho_{i3},
\end{eqnarray}
respectively. Here the $b_{K}=3/(8f_{\pi}^{\ast 2})\approx$0.333 GeVfm$^{3}$, the $a_{K}$ and $a_{\overline{K}}$ are 0.18 GeV$^{2}$fm$^{3}$ and 0.31 GeV$^{2}$fm$^{3}$, respectively, which result in the strengths of repulsive kaon-nucleon (KN) potential and of attractive antikaon-nucleon $\overline{K}$N potential with the values of 27.8 MeV and -100.3 MeV at saturation baryon density for isospin symmetric matter, respectively. The $\tau_{3}$=1 and -1 for the isospin pair K$^{+}$($\overline{K}^{0}$) and K$^{0}$(K$^{-}$), respectively. The parameters $c_{K}$=0.0298 GeV$^{2}$fm$^{3}$ and $d_{K}$=0.111 GeVfm$^{3}$ determine the isospin splitting of kaons in neutron-rich nuclear matter. The optical potential of kaon is derived from the in-medium energy as $V_{opt}(\textbf{p},\rho)=\omega(\textbf{p},\rho)-\sqrt{\textbf{p}^{2}+m_{K}^{2}}$. The values of $m^{\ast}_{K}/m_{K}$=1.056 and $m^{\ast}_{\overline{K}}/m_{\overline{K}}$=0.797 at normal baryon density are concluded with the parameters in isospin symmetric nuclear matter. The effective mass is used to evaluate the threshold energy for kaon and antikaon production, e.g., the threshold energy in the pion-baryon collisions $\sqrt{s_{th}}=m^{\ast}_{Y} + m^{\ast}_{K}$.

\begin{figure*}
\includegraphics{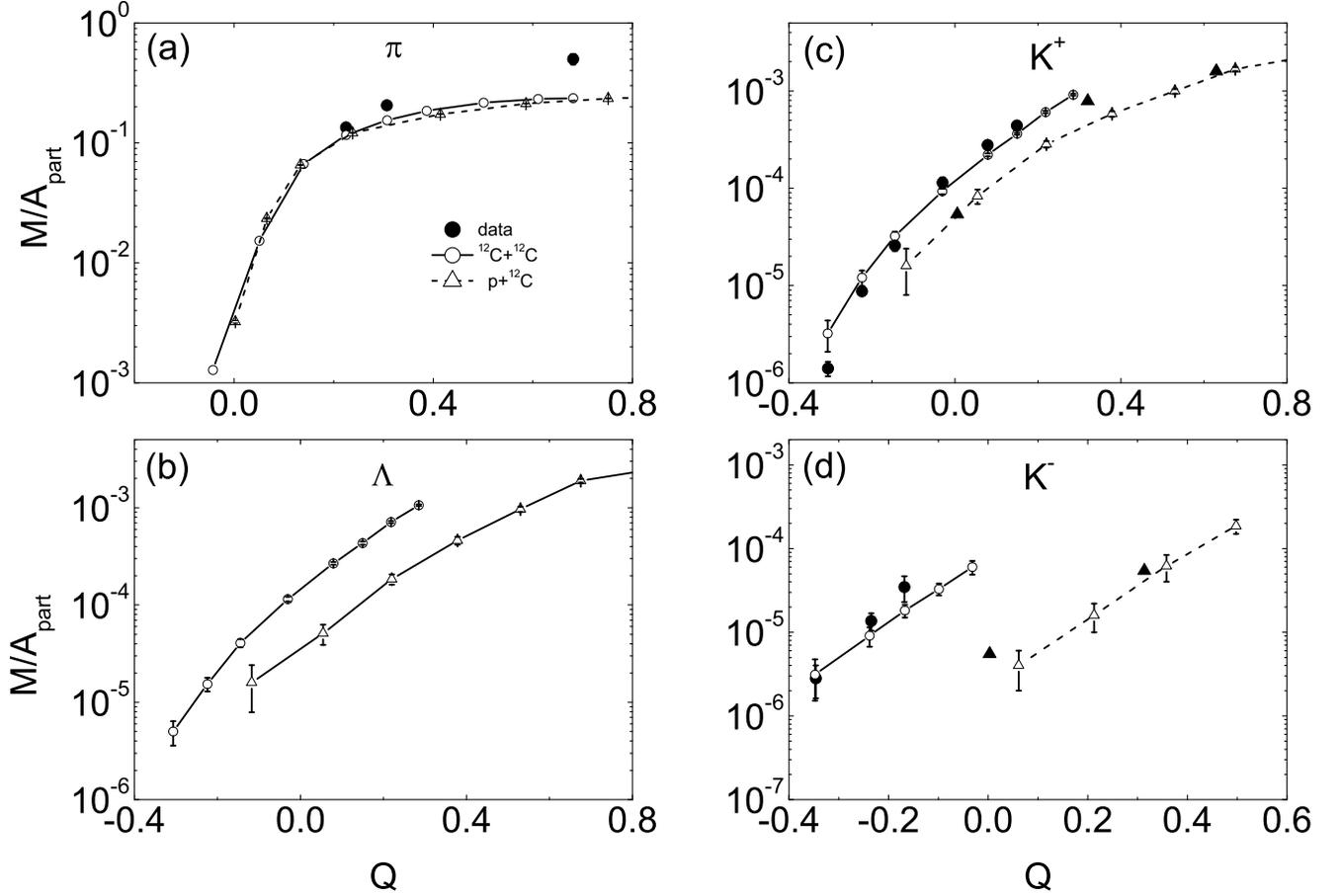}
\caption{\label{fig:wide} Multiplicities of $\pi$, $\Lambda$, K$^{+}$ and K$^{-}$ per participating nucleon in the reactions of $^{12}$C+$^{12}$C and p+$^{12}$C as a function of the available energy, i.e., the $\emph{Q}$ value being the difference of invariant energy to the threshold value $Q=\sqrt{s}-\sqrt{s_{th}}$.}
\end{figure*}

In the LQMD model, we have included all the coupled channels in producing pions, hyperons, kaons and antikaons near threshold energies. From the previous studies in heavy-ion collisions, it has been shown that the kaon (antikaon)-nucleon potential plays a significant role on the strangeness production and dynamical emission in phase space, which reduces (enhances) the kaon (antikaon) yields and is pronounced when the incident energy is close to the threshold value of kaon (antikaon) production \cite{Fe13}. We computed the production of $\pi$, $\Lambda$, K$^{+}$ and K$^{-}$ per participating nucleon in collisions of protons on $^{12}$C and in the $^{12}$C+$^{12}$C reaction as shown in Fig. 1. The full symbols are the experimental data for pion production from TAPS collaboration \cite{Av03} and for K$^{+}$ (K$^{-}$) measurements from KaoS collaboration \cite{La99,Sc06}. It is obvious that strange particles could be more easily produced in heavy-ion collisions, in particular for the production of K$^{-}$. The enhancement is caused from the compression of colliding nuclei and strangeness exchange reactions, i.e., the channels of $\pi Y\rightarrow$K$^{-}$N and $YN\rightarrow$K$^{-}$NN. The in-medium modifications on the pion production are not considered in the present model, which would play significant roles on the pion dynamics near threshold energies. Once pions are produced, subsequent reabsorption reactions via collisions of pions and surrounding nucleons take place. Thus, the similar excitation functions for pion production in the reactions of p+$^{12}$C and $^{12}$C+$^{12}$C are concluded.

\begin{figure*}
\includegraphics{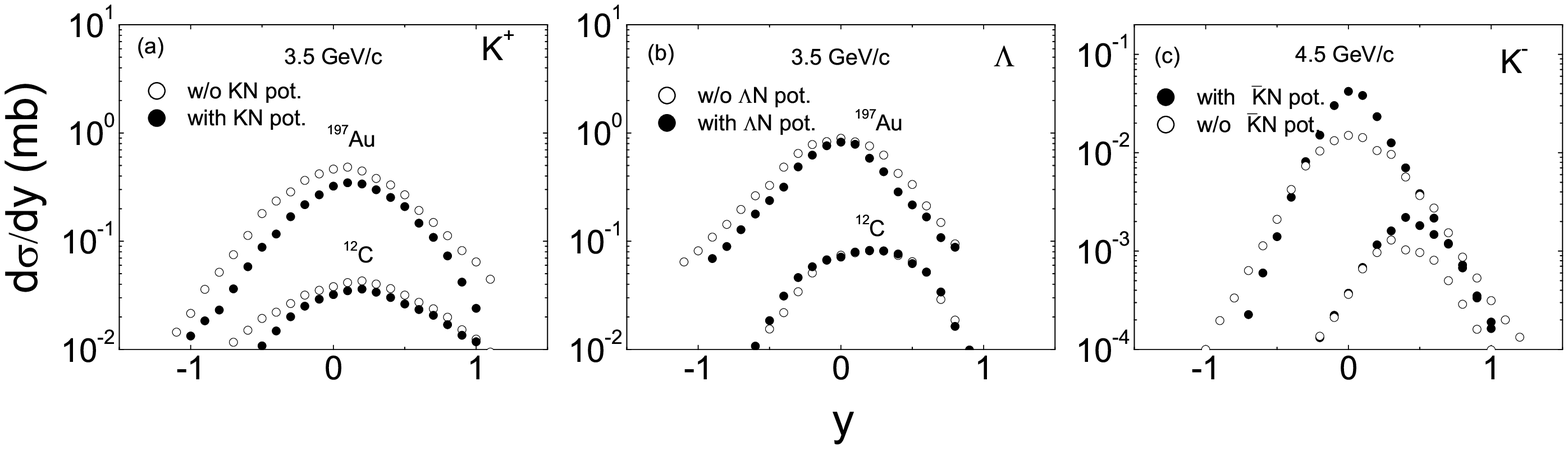}
\caption{\label{fig:wide} Rapidity distributions of strange particles produced in collisions of proton on $^{12}$C and $^{197}$Au.}
\end{figure*}

\begin{figure*}
\includegraphics{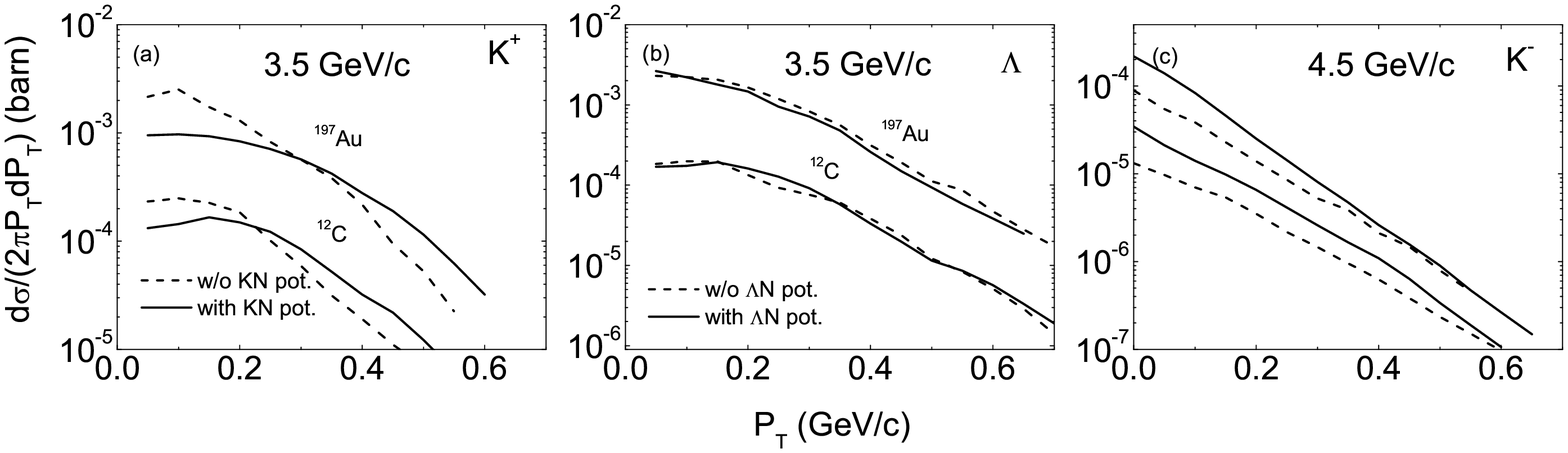}
\caption{\label{fig:wide} The same as in Fig. 2, but for the transverse mass spectra.}
\end{figure*}

The phase-space distribution of strange particles produced in heavy-ion collisions was modified in nuclear medium in comparison to in-vacuum case \cite{Fe13}. In this work, the in-medium effect on strangeness dynamics in proton induced reactions is to be investigated. Shown in Fig. 2 is the longitudinal rapidity distributions of K$^{+}$, $\Lambda$ and K$^{-}$ in collisions of proton on $^{12}$C and $^{197}$Au with and without inclusion of the mean-field potentials in nuclear medium. The kaon-nucleon and antikaon-nucleon potentials change the structure, in particular for the heavy target $^{197}$Au. Particles are produced towards forward emissions, in particular for lighter target. The transverse momentum spectra of the strange particles are also computed as shown in Fig. 3. It is pronounced that the kaon-nucleon potential reduces the low-momentum K$^{+}$ production and enhances the tail part of the spectra. However, the antikaon-nucleon potential leads to an opposite contribution on the K$^{-}$ emission because of the attractive interaction with surrounding nucleons. However, influence of the lambda-nucleon potential on $\Lambda$ dynamics in proton-nucleus collisions is negligible.

\begin{figure*}
\includegraphics{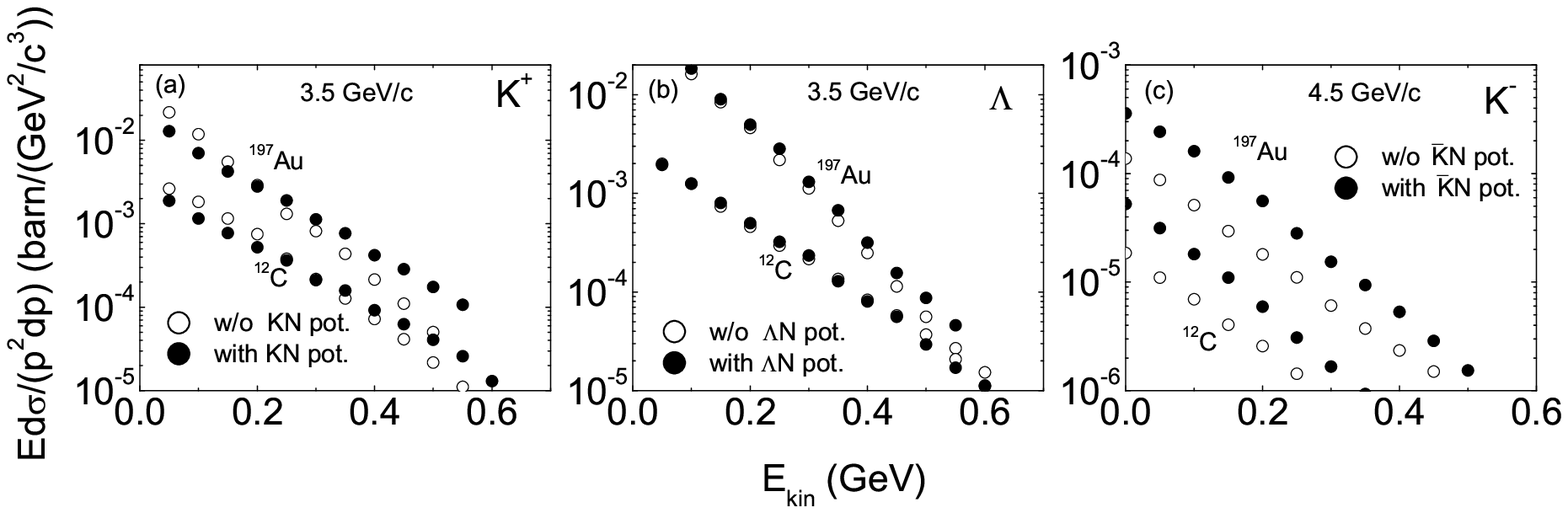}
\caption{\label{fig:wide} The inclusive spectra of strange particles (K$^{+}$, $\Lambda$ and K$^{-}$) produced in the proton induced reactions on $^{12}$C and $^{197}$Au with and without inclusions of the in-medium potentials at incident momenta of 3.5 GeV/c and 5 GeV/c, respectively.}
\end{figure*}

The interaction potential of strange particle and nucleon is of significance in the formation of hypernucleus in heavy-ion and proton-nucleus collisions, core structure of neutron star etc. However, it is not well understood up to now, in particular in the dense nuclear matter. The proton-nucleus collisions have advantage in constraining the in-medium properties of strangeness around the normal baryon density. Shown in Fig. 4 is the kinetic energy spectra of invariant cross sections of strange particles produced in collisions of proton on $^{12}$C and $^{197}$Au. It is pronounced that the kaon spectra become more flat with inclusion of the kaon-nucleon potential. However, the $\overline{K}$N potential exhibits an opposite contribution on antikaon dynamics. The $\Lambda$ hyperon production weakly depends on the in-medium potential. The effect is more obvious in heavy target. The conclusions are similar to heavy-ion collisions \cite{Fe13}.

\begin{figure*}
\includegraphics{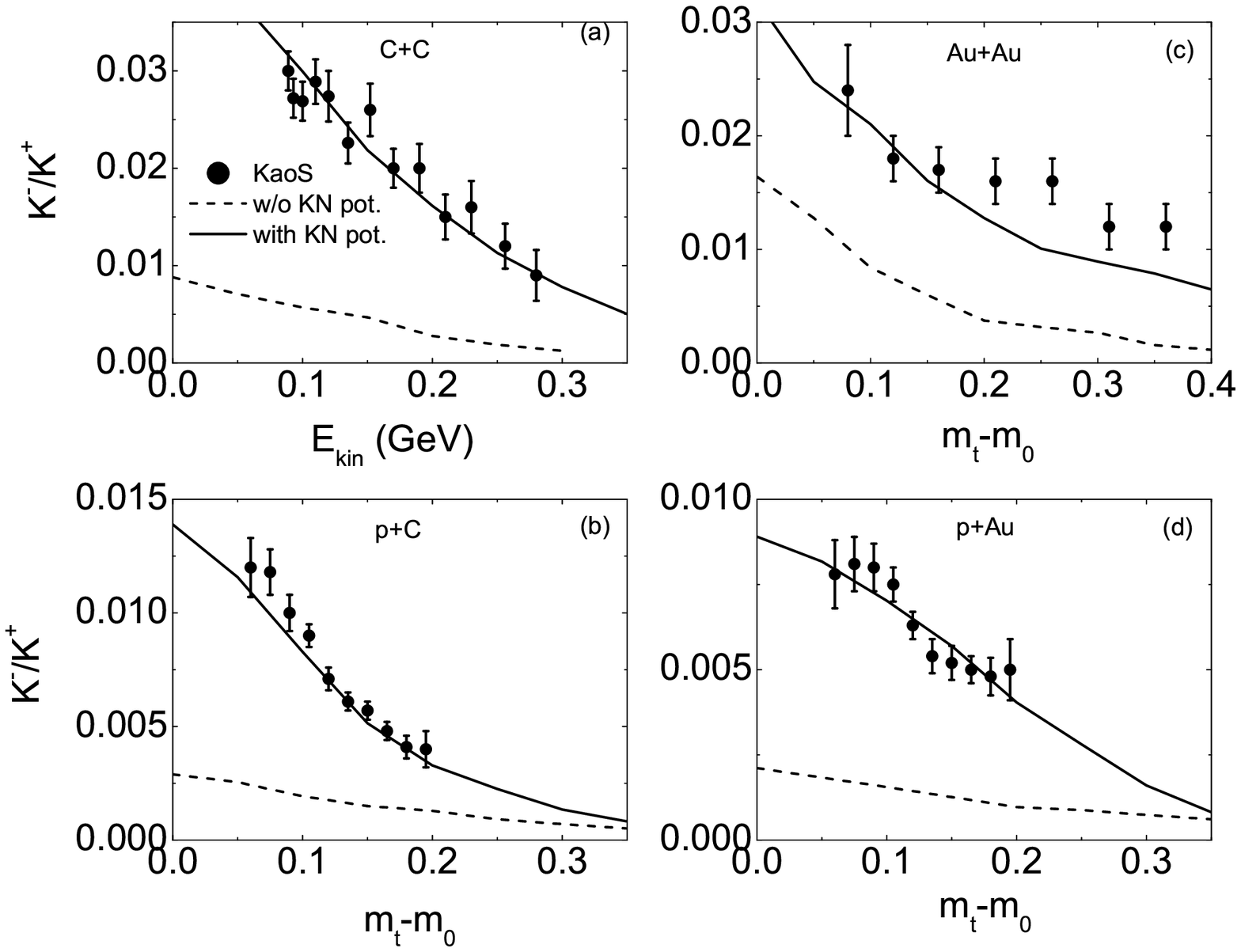}
\caption{\label{fig:wide} Ratio of K$^{-}$/K$^{+}$ as a function of transverse mass (kinetic energy) in collisions of $^{12}$C+$^{12}$C and protons on $^{12}$C at the beam energies of 1.8\emph{A} GeV and 2.5 GeV, respectively.}
\end{figure*}

More pronounced information of the in-medium effects has been investigated from the K$^{-}$/K$^{+}$ spectrum in heavy-ion collisions. A deeply attractive $K^{-}N$ potential being the value of -110$\pm$15 MeV was obtained at saturation density and weakly repulsive $K^{+}N$ potential has been concluded to be 25$\pm$10 MeV from heavy-ion collisions\cite{Li97,Ca99,Fu06,Ha12}. The in-medium modifications influence the kaon production and dynamical emissions in phase space, i.e., inclusive spectrum, collective flows etc. We compared the effects of kaon(antikaon)-nucleon potentials in the $^{12}$C+$^{12}$C reaction as a function of kinetic energy and in collisions of $^{197}$Au+$^{197}$Au and proton on $^{12}$C and $^{197}$Au versus the transverse mass ($m_{t}=\sqrt{p_{t}^{2}+m_{0}^{2}}$ with $p_{t}$ being the transverse momentum and the mass of kaon (antikaon) $m_{0}$) as shown in Fig. 5. The experimental data from KaoS collaboration \cite{La99,Fo03,Sc06} can be nicely reproduced within inclusion of the mean-field potentials for kaons and antikaons, in which a weakly repulsive $KN$ potential of the order of 28 MeV and a deeply attractive $\overline{K}N$ potential of -100 MeV at normal nuclear density are used in the LQMD model. The attractive $\overline{K}N$ potential reduces the threshold energies associated with enhancing $K^{-}$ production. However, the $KN$ potential leads to an opposite contribution for the $K^{+}$ emission. The value of $K^{-}/K^{+}$ ratio is more sensitive to the $\overline{K}N$ potential because of the larger strength.

In summary, strangeness production in proton induced reactions at beam energies close to the thresholds has been investigated within the LQMD transport model. The strange particles are strongly suppressed in proton-nucleus collisions in comparison with  heavy-ion collisions. A weakly repulsive $KN$ potential being 28 MeV and the attractive $\overline{K}N$ potential being -100 MeV in nuclear medium are concluded in comparison to the KaoS data for heavy-ion and proton-nucleus collisions. The mean-field potentials lead to a more flat and steeper spectra of the transverse momentum distributions and inclusive cross sections for $K^{+}$ and $K^{-}$ production, respectively.

This work was supported by the Major State Basic Research Development Program in China (No. 2014CB845405 and 2015CB856903), the National Natural Science Foundation of China Projects (Nos 11175218 and U1332207) and the Youth Innovation Promotion Association of Chinese Academy of Sciences.


\begin{thebibliography}{99}

\bibitem{Gi95} B. E. Gibson and E. V. Hungerford III, Phys. Rep. \textbf{257}, 349 (1995).
\bibitem{Fr07} E. Friedman and A. Gal, Phys. Rep. \textbf{452}, 89 (2007).
\bibitem{Ba97} R. Barth \emph{et al.} (KaoS Collaboration), Phys. Rev. Lett. \textbf{78}, 4007 (1997).
\bibitem{Me00} M. Menzel \emph{et al.} (KaoS Collaboration), Phys. Lett. B \textbf{495}, 26 (2000).
\bibitem{Fo03} A. F\"{o}rster \emph{et al.} (KaoS Collaboration), Phys. Rev. Lett. \textbf{91}, 152301 (2003).
\bibitem{Li97} G. Q. Li, C. H. Lee, and G. E. Brown, Nucl. Phys. A \textbf{625}, 372 (1997); G. Q. Li and G. E. Brown \textbf{636}, 487 (1998).
\bibitem{Ca99} W. Cassing and E. L. Bratkovskaya, Phys. Rep. \textbf{308}, 65 (1999).
\bibitem{Fu06} C. Fuchs, Prog. Part. Nucl. Phys. \textbf{56}, 1 (2006).
\bibitem{Ha12} C. Hartnack, H. Oeschler, Y. Leifels, E. Bratkovskaya, and J. Aichelin, Phys. Rep. \textbf{510}, 119 (2012).
\bibitem{Fe11} Z. Q. Feng, Phys. Rev. C \textbf{83}, 067604 (2011); \textbf{84}, 024610 (2011); \textbf{85}, 014604 (2012); Nucl. Phys. A \textbf{878}, 3 (2012); Phys. Lett. B \textbf{707}, 83 (2012).
\bibitem{Fe13} Z. Q. Feng, Nucl. Phys. A \textbf{919}, 32 (2013); Phys. Rev. C \textbf{87}, 064605 (2013); Z. Q. Feng, H. Lenske, \emph{ibid.} \textbf{89}, 044617 (2014).
\bibitem{Ka86} D. B. Kaplan and A. E. Nelson, Phys. Lett. B \textbf{175}, 57 (1986).
\bibitem{Sc97} J. Schaffner-Bielich, I.N. Mishustin, and J. Bondorf, Nucl. Phys. A \textbf{625}, 325 (1997).
\bibitem{Av03} R. Averbeck, R. Holzmann, V. Metag, and R. S. Simon, Phys. Rev. C \textbf{67} (2003) 024903.
\bibitem{La99} F. Laue \emph{et al.}, Phys. Rev. Lett. \textbf{82}, 1640 (1999).
\bibitem{Sc06} W. Scheinast \emph{et al.}, Phys. Rev. Lett. \textbf{96}, 072301 (2006).

\end{thebibliography}
\end{document}